\documentclass[a4paper,11pt]{article}
\pdfoutput=1 % if your are submitting a pdflatex (i.e. if you have
\usepackage{jheppub} % for details on the use of the package, please
\usepackage[T1]{fontenc} % if needed

\newcommand{\Frac}[2]{\displaystyle\frac{#1}{#2}}
\newcommand{\EV}{\text{eV}}
\newcommand{\KeV}{\text{KeV}}
\newcommand{\MeV}{\text{MeV}}
\newcommand{\GeV}{\text{GeV}}
\newcommand{\TeV}{\text{TeV}}

\title{\boldmath  Application of Flavor Moonshine}

\author{H. Sugawara}

\affiliation{KEK,\\ Oho, 1 - 1, Tsukuba City, 305-0801 Ibaraki Japan}

\emailAdd{sugawara@post.kek.jp}

\abstract{
In our previous paper \cite{a}, we presented the application of Moonshine to
the flavor physics based on the work of Cohn and Deutsch and Bagger and
Wess \cite{b}, J. Conway and S. Norton \cite{c} and T.Eguchi, H.Ooguri and
Y.Tachikawa \cite{e} laid down the mathematical basics.
We present here the extension of our previous paper \cite{a} by including the
supersymmetry in addition to the modular symmetry \cite{a}, \cite{e}. 
We calculate all the lepton masses and the quark masses and their super
partner masses by using the one input particle mass for each family.
The two valued modular function is defined in equation (\ref{eq3-7}).  
We define the two valued modular function of degree K in equation
(\ref{eq3-2}).
We use $K = 1$ two valued modular function for the charged leptons and 
$K = 2$ two valued modular function for the neutrino and for the up quarks,
and $K = 3$ two valued modular function for the down quarks, where $2K$ is
the level of the two valued modular function.

We emphasize that we have three particle masses which can be measured by
the existing accelerators such as KEK B factory or CERN LHC.
One is super electron mass ``$M_{e,s} =24.73 \MeV$'', which is 
approximately 25 times that of electron.
The Super $d$, $s$ and $b$ quark masses are $0 \GeV$, $89.165 \GeV$, 
and $2964.7 \GeV$.
Super down quark mass is 0 and super strange quark mass can be measured
both in KEKB and in CERN LHC.
The super bottom quark mass can be measured only in LHC.

}

\begin{document} 
\maketitle
\flushbottom
%%%%%%%%%
\section{Introduction}
\label{sec:intro}
Followings table show how to measure the masses of super partners of
leptons and quarks in KEK or in CERN.

The value of masses is given, for example, in the table
% \begin{table}[hbp]
% \begin{table}[!ht]
\begin{gather*}
\begin{array}{|c|c|c|}
\hline
M_{e,s} & M_{\mu,s} & M_{\tau,s} 
\\ \hline 
 24.73 \MeV &
 4.03 \times 10^{12} \MeV &
 8.46 \times 10^9 \MeV
\\ \hline 
\end{array}
\end{gather*}
% \caption{\label{tab:1} 
% }
% \end{table}

All the other masses are given later in this paper.

\noindent\textbf{KEKB} \\
The following table shows that KEKB can measure the super electron mass
and the largest and the next largest neutrino masses. 
KEKB can also measure the super up quark mass.
% \begin{table}[hbp]
%\begin{table}[!ht]
\begin{gather*}
\begin{array}{|c|c|c|c|}
\hline
M_{e,s} & M_{2 \nu, s} & M_{3 \nu, s} & M_{u,s} 
\\ \hline
24.73 \MeV &
0.973 \EV &
1.7835 \times 10^4 \EV &
25.7 \MeV 
\\ \hline 
\end{array}
\end{gather*}
% \end{table}

KEKB II has the luminocity = $2.11 \times 10^{34} cm^{-2} s^{-1}$.
Event number of ($e e^+ \rightarrow \text{two spin 0 bosons}$) 
is $4.0 \times  10^4 /\text{sec}$ (appendix):
% \begin{table}[hbp]
\begin{table}[!ht]
\begin{gather*}
\renewcommand{\arraystretch}{2.0}
\begin{array}{|c|}\hline
\sigma(e e^+ \rightarrow e e^+) 
= \Frac{0.1068}{11^2 \times 0.1^2} = 0.088 \, (\MeV)^{-2}
\\ \hline
\begin{split}
\Frac{\sigma(e e^+ \rightarrow \text{two ~spin~0~bosons})}{
      \sigma(e e^+ \rightarrow e e^+)}
& 
=  \Frac{3.14}{3 \times 137 \times 121000 \, (\MeV)^2}
\\ & 
=  6.31 \times 10^{-8} \, (\MeV)^{-2}
\end{split}
\\ \hline 
\sigma (e^- e^+ \rightarrow \tilde{e}^{-} + \tilde{e}^{+}) = \Frac{\pi \alpha^2}{3 s} \Bigl(1 - \Frac{4 m^2}{s}\Bigr)^{3/2}
\sim 7.17 \times 10^{-7} \,(\MeV)^{-2}
\\ \hline 
\end{array}
\end{gather*}
\caption{\label{tab:2}
Event number of ($e e^+ \rightarrow \text{ two spin 0 bosons}$)
} 
\end{table}

The above table shows that the cross section to produce super electron
is seven orders of magnitude smaller than that of $e^+e^-$ elastic
scattering.

The ratio of the cross section to produce super neutrino to the 
$\sigma(e e^{+} \rightarrow ee^{+})$ is given by:
\begin{equation}
\Frac{\Frac{1.17}{M_W^2}\times 10^{-5}}{\alpha}
=
\Frac{\Frac{1.17}{80.38^2}\times 10^{-5}}{\Frac{1}{137}}
= 2.4 \times 10^{-7}
\end{equation}
For a nucleus of charge $Z$, the nonrelativistic binding energy scale
as:
\begin{equation}
E_n = - \frac{\mu c^2 (Z \alpha)^2}{2 n^2}
\end{equation}
Super nucleus emmits proper frequency $\gamma$ ray :
\begin{equation}
E(1s) \simeq 48.4 \times 13.6 \EV \simeq 658 \EV
\end{equation}
\begin{equation}
E(\text{gamma}) = 0.494 \KeV
\end{equation}
The mass of the super electron using the above two gamma rays created in
the beam dump.
The cross is about $10^{-3}$ times smaller than that of the
Bhabha scattering. (see appendix) :

$\mu$-on pair production $e^+ e^- \rightarrow \mu^+ \mu^-$
\begin{equation}
\frac{d \sigma}{d \Omega} \propto 
1 + \cos^2 \theta + (1 - \beta_\mu^2) \sin^2 \theta
\end{equation}
Scalar super-electron pair production 
$e^+ e^- \rightarrow \tilde{e} + \tilde{e}$
\begin{equation}
\frac{d \sigma}{d \Omega} \propto 
\beta^3_{\tilde{e}} \sin^2 \theta
\end{equation}

\noindent
\textbf{CERN LHC}\\
The following table shows that LHC can measure the super electron mass,
heaviest and the next heaviest neutrino masses, super muon mass, super 
strange quark mass and super bottom quark mass. 

% \begin{table}[hbp]
\begin{table}[!ht]
\begin{gather*}
\begin{array}{|c|c|c|}
\hline
M_{e,s} & M_{\mu, s} & M_{2 \nu, s}
\\ \hline
24.73 \MeV & 1.17 \times 10^6 \MeV & 0.973 \EV
\\ \hline
M_{3 \nu,s} & M_{s, s} & M_{b, s}
\\ \hline
1.7835 \times 10^4 \EV & 
89.165 \GeV & 
2964.7 \GeV
\\ \hline
\end{array}
\\
\begin{array}{|l|}
\hline
\displaystyle
\sigma(pp \rightarrow s+X) 
= \sum_{ij} \int_0^1 x_1\,dx_1 \int_0^1 x_2\,dx_2\, 
   f_i(x_1,\mu_F)\, f_j(x_2,\mu_F)\, \sigma_{ij}
\\ \hline 
\end{array}
\end{gather*}
\caption{\label{tab:4} 
Inclusive cross section for proton - proton $\rightarrow$ scaler +
anything.
$f_i (x_1, \mu_F)$ is the $i$-th parton (scalars) function. 
The results depend on the values of $ f_i (x_1, \mu_F)$.
$\sigma_{i j}$ is the cross section of the $i$-the and $j$-th
quark (partons).}
\end{table}

%%%%%%%%%
\section{Purpose of this work}

Our motivation, since last work (paper \cite{a}), is to understand the
extension of symmetry so that we eventually reach the largest symmetry
which enables us to calculate all the masses with a single free parameter 
(to be published).
%particle masses without any parameter.
Supersymmetry is to put fermions and bosons in a single representation
\cite{f} and \cite{g}. 
As mentioned in the abstract, we have supersymmetry in addition to
modular symmetry of all the fermions using electron mass, first and
second highest neutrino mass ratio, $t$-quark and $b$-quark masses as the 
inputs.

%The next step is to all the families in a single multiplet. 
The necessary symmetry is already available: grand unification.
The last thing we should do, in future, is to find symmetry to include
all the gauge particles, and graviton.
In the next section we write Lagrange formalism of particle masses,
loosely dependent on reference \cite{c}. 
We write the Lagrangean for each family. 
The family is identified by parameter $K$ two times of which is the
level of two variable modular function. 
We add see-saw Lagrangean to the one in the discussion section of
neutrino mass.
The vacuum value of Higgs field $\langle A_H \rangle$ spontaneously
breaks the supersymmetry of the $L_K$.
We briefly discuss two valued modular invariance.
Two valued modular invariance is explicitly broken by $< A_H >$. 
%`$\Phi^{+}_i \Phi_i$'.
The fields $\Phi_i$ and $\Phi_H$ are functions of super space parameter
$x$, $\theta$ and $\overline{\theta}$.

%%%%%%%%%
\section{Lagrangean formalism of particle masses}

We start from $L=\sum_{K=1}^3L_K$, where $K$ is defined in 
equation (\ref{eq3-7}).
$2K$ is the defined as the level of the two valued
``Modular Transformation''.
\begin{gather} \label{eq3-1}
\begin{split}
-L_K &
= \Phi^+_i \Phi_i \Phi^+_H \Phi_H |_1
+ \mu^2 \Phi^+_i \Phi_i |_{\overline{\theta}\overline{\theta} \theta \theta}
+ \mu^2 \Phi^+_H \Phi_H |_{\overline{\theta}\overline{\theta} \theta \theta}
\\ &
+ \frac{\mu \mu_K^{-1}}{3} \sum \left(g g^t \right)_{i j}^{1/2}
   \left[\Bigl(\Phi_H \Phi^{*}_i \Phi_j \Bigr)\Bigr|_{\overline{\theta} \theta} 
+ \alpha \mu \Phi_H \Phi_i^{*} \Phi_j\right]
% \\ &
+ \lambda\left[\Phi^{\ast}_H \Phi^{\ast}_H \Phi_H \Phi_H \Bigr|_1\right] 
+ \text{h.c.}
\end{split}
\end{gather}
``$g$'' is the coefficient matrix two valued modular function for each
family such as in equation (\ref{eq14}).
In the case of neutrino family, we use the following see-saw
Lagrangean :
\begin{gather} \label{eq3-2}
\begin{split}
&
L_{\text{see-saw mass}} 
\\ & \quad
% = \frac{1}{2} \sum (g g)_{i j}^t   (???)
= \frac{1}{2} \sum (g g^t)_{i j}
\Bigl[
    \mu_\nu A^{*}(x)_{L, i\nu} 
      + \sqrt{2} \mu_\nu^{1/2} \overline{\psi}_{L, i\nu}(x) \theta, \;
    \mu_\nu A(x)_{R, j\nu} 
      + \sqrt{2} \mu_\nu^{1/2} \overline{\theta} \psi_{R, j\nu}(x)
\Bigr]
\\ & \quad\quad
\begin{pmatrix} 1 & 0 \\ 0 & 1 \end{pmatrix}
\Bigl[
    \mu_\nu A^{*}(x)_{R, i\nu} 
      + \mu_\nu^{1/2} \overline{\psi}_{R, i\nu}(x) \theta, \;
    \mu_\nu A(x)_{L, j\nu} 
      + \sqrt{2} \mu_\nu^{1/2} \overline{\theta} \psi_{L, j\nu}(x)
\Bigr]
+ \text{h.c.}
\end{split}
\end{gather}
where following is usual anticommuting algebra :
\begin{gather}
[\theta, \theta ]_+ = 0. 
\end{gather}
$\Phi_i$ is $i$-th component of a family. 
$\Phi_H $ is the Higgs field which appears in Lagrangean $L_K$.
Matrix $g_{i j}$ is the Taylor expansion coefficient matrix of two
variable modular function:
\begin{equation}
J(q, r) = \sum_{i, j=-\infty}^\infty g_{i j}[K] \, q^i r^j
\end{equation}
\begin{equation}
\Phi_i 
= A_i(y^\mu, z, z') 
  + \sqrt{2}\overline{\theta} \psi_i(y^\mu, z, z')
    \rightarrow \Phi_i(\alpha z + \beta, \alpha' z' + \beta')^{-2 K}
\end{equation}
\begin{equation}
\Phi_H 
= \mu_H A_H(y^\mu, z, z') 
  + \sqrt{2} \mu_H^{1/2} \overline{\theta} \psi_H(y^\mu, z, z')
\label{eq7}
\end{equation}
\begin{equation}
J_K(z, z') \rightarrow
J_K(z, z') \left(\frac{\alpha z + \beta}{\gamma z + \delta},
  \frac{\alpha' z' + \beta'}{\gamma' z' + \delta'}\right)^{2 K}
\label{eq3-7}
\end{equation}
`$\rightarrow$' means the modular transformation.
The single valued modular function is defined :
\begin{gather}
G_{2K} = A_{2K}+B_{2K}\sum q^b r^a s_{2K-1}(a+\sqrt 2 b)
,\qquad
K \geq l
\end{gather}
where ``$G_{2K}$'' is single variable modular transformation 
(reference  \cite{b}). 
$2K$ is defined to be it’s level with,
\begin{gather}
(A_2, B_2) = (1,48)
,\quad 
(A_4, B_4) = (11,480)
, \quad
(A_6, B_6) = (361,1008)
\\
S_u(\alpha) = \sum |\mu \mu'|^u(\mu)(\alpha)
\end{gather}
with $\mu$ summed over all ideals $\mu$ dividing $\alpha$.
In the present paper, we use two valued modular function unlike in reference [1].

%%%%%%%%%
\section{Leptons and quarks and their super partners}

We define $r$ and $q$ as follows:
\begin{gather}
r = e^{2\pi i z}
, \qquad
q = e^{2\pi i z'}
\\
g_{i j} = \int \int J(r,q) r^{-i} q^{-j} \; dz \; dz'
\end{gather}

\begin{itemize}
\item[(a)]
We have, for $K = 1$ (lepton) :
\begin{gather}
g_{i j} =
\begin{pmatrix}
  1 &   0 &   0 \\
144 &  48 &   0 \\
720 & 384 & 336
\end{pmatrix} 
\end{gather}

\item[(b)]
We have, for $K = 2$ (neutrino) :
\begin{gather} \label{eq14}
g_{i j}^2 =
\begin{pmatrix}
  1 &   0 &   0 \\
144 &  48 &   0 \\
720 & 384 & 336 \\
\end{pmatrix} 
\begin{pmatrix}
  1 &   0 &   0 \\
144 &  48 &   0 \\
720 & 384 & 336 \\
\end{pmatrix} 
\end{gather}

\item[(c)]
We have $\Bigl(\Frac{+2}{3} \text{quark}\Bigr)$ for $K = 2$:
\begin{gather} \label{eq15}
G_4=
\begin{pmatrix}
    11 &      0 &     0 \\
  4320 &    480 &     0 \\
280800 & 165120 & 35040 \\
\end{pmatrix}
\end{gather}

\item[(d)]
We have $\Bigl(\Frac{-1}{3} \text{quark}\Bigr)$ for $K = 3$:
\begin{gather}
H_6=
\begin{pmatrix}
 0 &   0 &  0 \\
 1 &   0 &  0 \\
12 & -16 & -2 \\
\end{pmatrix} 
\end{gather}
\end{itemize}
Modular transformation is valid in the case of charged lepton sector, 
charge $+\frac{2}{3}$ sector and charge $-\frac{1}{3}$ sector.
It is broken in neutrino case.

%%%%%%%%%
\section{Vacuum value, $\Phi_H$ and the Lagrangean in terms of component fields}

Higgs field is given by:
\begin{gather}
\begin{split}
\Phi_H &
= A_H(x) 
+ \sqrt{2} \mu^{-1/2} \overline{\theta} \psi_H(x)
+ i \mu \overline{\theta} \theta \gamma_\mu \partial_\mu A_H(x)
+ 2 \theta \theta \overline{\theta} \overline{\theta} A_H(x)
\\ & \quad
+ \sqrt{2} \mu^{-1/2} \overline{\theta} 
  \Bigl(\psi_H(x) 
       + i \mu^{-1} \overline{\theta} \theta \gamma_\mu \partial_\mu \psi_H(x)
  \Bigr)
\end{split}
\end{gather}
We get Higgs Lagrangean from equation (\ref{eq3-1}):
\begin{align}
\begin{split}
-L_H &
= \mu^2 A^*(x) A(x) 
- \lambda A^*(x) A^*(x) A(x) A(x)
+ \overline{\psi}_H(x) \gamma_\mu \partial_\mu \psi_H(x)
\\ & \quad
+ \frac{1}{\mu_K} A(x) \overline{\psi}_H(x) \psi_H(x)
+ \text{h.c.}
\end{split}
\end{align}
Higgs potential is given by:
\begin{gather}
V_{H} 
= - \mu^2 A^{\ast}(x) A(x) 
  + \lambda A^{\ast}(x) A^{\ast}(x) A(x) A(x)
\end{gather}
Vacuum value is obtained from:
\begin{gather}
\frac{\partial V_{H}}{\partial (A^*(x) A(x))} 
= - \mu^2 
  + 2 \lambda A^{\ast}(x) A(x) = 0
\end{gather}
We get
\begin{gather}
A = \frac{1}{\sqrt{2 \lambda}} \mu
\\ 
M_H^2 = - \mu^2 + 2 \lambda \mu^2 = 2 \mu^2
%\qquad
%\lambda
% ,\\
% M_H^2 &= 2 \mu^2
\label{eq5-6}
\end{gather}
$\lambda$ is taken to be 3/2  without loss of generality.
This gives:
\begin{gather}
\mu = 88.642 \GeV
\end{gather}
Here we used:
\begin{gather}
M_H = 125.11 \GeV
\end{gather}
We get Higgsino mass from Lagrangean (\ref{eq3-1}) :
\begin{gather}
M_{H,s} = \frac{3}{2} \frac{\mu}{\mu_K}
    A(x) \overline{\psi}_H(x) \psi_H(x)
\end{gather}
We have the ratio of Higgsino mass \cite{h}, vs Higgs mass: 
$m_{H,s} / m_{H} =200/125.11=1.60$ and 
$m_{H,s} / m_{H} = 1100/125.11 =8.79$ \cite{i} 
with
$\lambda = 1.5$ and $\mu_K = 237.056$.
This value of the Higgsino mass is consistent with the Higgsino being
the dark matter.
Actually, the real experimental value of 4-Higgs :
\begin{equation}
-22.8 \lesssim \lambda \lesssim 23.9
\end{equation}

We used the input value (1.1TeV from \cite{h}):
\begin{equation}
% \mu_K = \frac{3 \times 88.642}{2 \times 1100} = 0.66 \quad\text{(???)}
% ,\\
\mu_K = \frac{3 \times 88.642}{2 \times 1100} = 0.12
\end{equation}
Here we cannot predict the Higgsino mass but obtained from the dark matter experiment \cite{h}, \cite{i}.

%%%%%%%%%
\section{Charged lepton masses}

\begin{gather}
\mu = 88.642 \MeV
\end{gather}
$\mu_K$ is obtained by electron mass as input.
\begin{gather}
\frac{3}{2} \mu \mu_l^{-1} = 1.744
\\
M_{f,l} 
= \frac{3}{2} \mu \mu_l^{-1} \sum_{i, j} \left(g g^t\right)^{1/2}_{i, j}
\\
M_{b,l} 
= \frac{3}{2} \mu^2 \mu_l^{-1} \sum_{i, j} \left(g g^t\right)
\end{gather}
Here we have:
\begin{gather}
\left(g g^t\right)_{i j}
=
\begin{pmatrix}
1 & 144 & 720 \\
144 & 23040 & 122112 \\
720 & 122112 & 778752 \\
\end{pmatrix}
\end{gather}
We get the following result using equation for electron mass as the input. 
We diagonalize the above matrix (\ref{eq35}).
After some numerical computation, we get:
\begin{gather}
\begin{pmatrix} 
m_e \\ m_\mu \\ m_\tau
\end{pmatrix}
= 1.742
\begin{pmatrix} 
0.293 & 0 & 0 \\
0 & 61.64 & 0 \\
0 & 0 & 893.31 \\
\end{pmatrix}
\end{gather}
The number $1.742$ is obtained using electron mass as the input 
(1-1 element of matrix (\ref{eq38})):
For the super charged lepton mass we have (from equation (\ref{eq32})):
\begin{gather} \label{eq32}
M_{b,l}^2
=
\begin{pmatrix} 
0.08 & 0 & 0 \\
0    & 3.8 \times 10^3 & 0 \\
0    &   0  & 7.98 \times 10^5 \\
\end{pmatrix} 
=
\begin{pmatrix} 
0.08 & 0 & 0 \\
0    & 3.8 \times 10^3 & 0 \\
0    &   0  & 7.98 \times 10^5 \\
\end{pmatrix} 
\end{gather}
The following table shows the charged lepton mass 
and the corresponding super charged lepton mass, electron mass being input.

\begin{table}[htb]
\begin{gather*}
\begin{array}{|c|c|c|}
\hline
M_e    = 0.511 & 
M_\mu  = 107.5 \MeV & 
M_\tau = 1558.3 \MeV 
\\ \hline
M_{e,s}   = 24.73 \MeV & 
M_{\mu,s} = 1.17 \times 10^6 \MeV &
M_{\tau,s}   = 2.47 \times 10^8 \MeV 
\\ \hline
\end{array}
\end{gather*}
\caption{electron and super electron mass}
\end{table}

%%%%%%%%%
\section{Charge $-1/3$ quark masses}

We use the mass formulae obtained from the Lagrangean (\ref{eq3-1}). 
$\mu$ is obtained using Higgs mass as in equation (\ref{eq5-6}). 
$\mu_K$ is obtained by using the electron mass as input.

%%%%%%%%%
\section{Sea-saw masses}

We start from the following See-saw mass Lagrangean:
\begin{gather}
\begin{split}
&
L_{\text{see-saw mass}} 
\\ & \quad
% = \frac{1}{2} \sum \left(g g\right)^t_{i j} ???
= \frac{1}{2} \sum \left(g g^t\right)_{i j}
  \Bigl[
      \mu_\nu A^*(x)_{L, i \nu} 
    + \sqrt{2} \mu_\nu^{1/2} \overline{\psi}_{L,i\nu}(x) \theta,\;
      \mu_\nu A(x)_{R, j \nu} 
    + \sqrt{2} \mu_\nu^{1/2} \overline{\theta} \psi_{R, j\nu}(x)
  \Bigr]
\\ & \quad\times
  \begin{pmatrix} 1 & 0 \\ 0 & 1 \end{pmatrix}
  \Bigl[
      \mu_\nu A^*(x)_{R, i \nu} 
    + \mu_\nu^{1/2} \overline{\psi}_{R,i\nu}(x) \theta,\;
      \mu_\nu A(x)_{L, j \nu} 
    + \sqrt{2} \mu_\nu^{1/2} \overline{\theta} \psi_{L, j\nu}(x)
  \Bigr]
  + \text{h.c.}
\end{split}
\end{gather}
We get the following neutrino and super neutrino masses:
\begin{gather} \label{eq34}
% M^2_{i j, \nu, b} = \mu_\nu^2 \left(g g\right)_{i j}^t    (???)
M^2_{i j, \nu, b} = \mu_\nu^2 \left(g g^t\right)_{i j}
\\  \label{eq35}
% M_{i j, \nu, f} = \mu_\nu \sqrt{\left(g g\right)_{i j}^t}  (???)
M_{i j, \nu, f} = \mu_\nu \sqrt{\left(g g^t\right)_{i j}}
\end{gather}
We have generically for the sea-saw mass,
\begin{gather}
\lambda^2 - (m_D + M_D) \lambda + (m_D M_D - m_M^2) = 0
\end{gather}
In our case sea-saw mass is equal to Dirac mass:
We solve the equation for large $m_M$.
We adopt the positive solution:
\begin{gather}
\lambda_+ = \frac{m_M^2}{m_D}
\end{gather}
We have as in equation (\ref{eq14}):
\begin{gather} \label{eq38}
g^2 =
\begin{pmatrix}
1 & 0 & 0 \\
144 & 48 & 0 \\
720 & 384 & 336 \\
\end{pmatrix}
\begin{pmatrix}
1 & 0 & 0 \\
144 & 48 & 0 \\
720 & 384 & 336 \\
\end{pmatrix}
\end{gather}
We get:
\begin{gather}
g^2(g^2)^t =
\begin{pmatrix}
1 & 7056 & 297936 \\
7056 & 55095552 & 2441975040 \\
297936 & 2441975040 & 123254638848 \\
\end{pmatrix}
\end{gather}
We solve the following equation of the determinant of the matrix of the
equation :
%\begin{gather} \label{eq40}
\begin{gather}
\text{Det}
\begin{pmatrix}
539137 - \lambda & 283392 & 112896 \\
91238544 &  1540.8 \times  10^7 - \lambda & 9772.8 \times  10^7 \\
505129680 & 8.737 \times  10^{10} & 
5.687 \times  10^{11} - \lambda \\
\end{pmatrix}
\label{eq8-8}
\end{gather}
We diagonalize the above matrix by solving:
\begin{gather}
\text{Det}
\begin{pmatrix}
1 - \lambda & 7056 & 297936 \\
7056 & 55095552 - \lambda & 2441975040 \\
297936 & 2441975040 & 123254638848 - \lambda \\
\end{pmatrix}
= 0
\label{eq8-9}
\end{gather}
We get:
\begin{gather}
\begin{pmatrix}
0.08 & 0 & 0 \\
0 & 6.711 \times  10^6 & 0 \\
0 & 0 & 1.23 \times  10^{11}
\end{pmatrix}
\label{eq8-10}
\end{gather}
%After some numerical computation, using equations (\ref{eq34}) and
%(\ref{eq35}) and the
After some numerical computation, using equations (\ref{eq8-9}) and
(\ref{eq8-10}) and the
highest neutrino mass
($5.1 \times 10^{-2} \EV$) as input, we get:

% \begin{table}[hbp]
\begin{table}[!ht]
\begin{gather*}
\begin{array}{|c|c|c|}
\hline
4.1  \times 10^{-8} \EV &
3.75 \times 10^{-4} \EV &
5.1  \times 10^{-2} \EV 
\\ \hline
M_{1\nu,s} = 1.16 \times 10^{-8} \EV &
M_{2\nu,s} = 0.973 \EV &
M_{3\nu,s} = 1.7835 \times 10^4 \EV 
\\ \hline
\end{array}
\end{gather*}
\caption{
Table for neutrino and super neutrino masses
}
\end{table}

%%%%%%%%%
\section{Charge $+2/3$ quark masses}

We use the following mass formulae just as in the case of charged lepton
mass:
\begin{gather}
M_{f,i} 
= \frac{3}{2} \mu \mu_{+2/3}^{-1} \sum_{i, j} \sqrt{\left(g g^t\right)}
\\
M_{b,i}^2 
= \frac{3}{2} \mu^2 \mu_{+2/3}^{-1} \sum_{i, j} \left(g g^t\right)
\end{gather}
$\mu_K$ is calculated top quark mass as input.
We use the definition of $G_4$ defined in equation (\ref{eq15}):
\begin{gather}
g = G_4 =
\begin{pmatrix}
11 & 0 & 0 \\
4320 & 480 & 0 \\
280800 & 165120 & 35040 \\
\end{pmatrix}
\end{gather}
We get:
\begin{gather}
g g^t =
\begin{pmatrix}
11 & 0 & 0 \\
4320 & 480 & 0 \\
280800 & 165120 & 35040 \\
\end{pmatrix}
\begin{pmatrix}
11 & 4320 & 280800 \\
0 & 480 & 165120  \\
0 & 0 & 35040 \\
\end{pmatrix}
\end{gather}
After some numerical computation we get:
% \begin{table}[hbp]
\begin{table}[!ht]
\begin{gather*}
\begin{array}{|c|c|c|}
\hline
0.376 \MeV &
0.731 \GeV &
172.5 \GeV
\\  \hline
0.0257 \GeV &
9 \times 10^2  \TeV &
2.9  \times 10^7 \TeV
\\ \hline
\end{array}
\end{gather*}
\caption{
charge + 2/3 quark mass (first row) and the + 2/3 squark mass (second row)
}
\end{table}

%%%%%%%%%
\section{Bottom, strange and down quarks masses}

We start from the mass formulae for charge $-\frac{1}{3}$:
\begin{gather} \label{eq47}
M_{f, -1/3} = \frac{3}{2} \mu \mu_{-1/3} \sum_{i, j} \left(g g^t\right)^{1/2}
\\
M^2_{b, -1/3} = \frac{3}{2} \mu^2 \mu_{-1/3}^{-1} 
  \sum_{i, j} \left(g g^t\right)^{1/2} (89.15)^2
\end{gather}
Using the formulae given in equation (\ref{eq47}) and equation for 
$\mu_{-1/3}$, we get:
\begin{gather}
(H_6)^t =
\begin{pmatrix}
0 & 0 & 0 \\
1 & 0 & 0 \\
12 & -16 & -2 \\
\end{pmatrix}
\begin{pmatrix}
0 & 1 & 12 \\
0 & 0 & -16 \\
0 & 0 & -2 \\
\end{pmatrix}
=
\begin{pmatrix}
0 & 0 & 0 \\
0 & 1 & 12 \\
0 & 12 & 404 \\
\end{pmatrix}
\\
\mu_{-1/3} = \frac{20.12  \times 2 \times (89.15)^2}{3\times 4.18}
 = 25503.7
\end{gather}
We get masses for $- 1/3$ quarks and $-1/3$ squarks with bottom quark mass as input:
\begin{table}[htb]
\begin{gather}
\begin{array}{|c|c|c|}
\hline
0 \GeV & 0.126 \GeV & 4.18\GeV
\\ \hline
0 \GeV &
89.165 \GeV &
2964.70 \GeV 
\\ \hline
\end{array}
\end{gather}
\caption{charge -1/3 quark mass (first row) and the charge -1/3 squark mass (second row)}
\end{table}

%%%%%%%%%
\section{Conclusion}

We proved the supersymmetry of the Lagrangean (\ref{eq3-1}) assuming that Higgs
field and its super partner are invariant under the supersymmetry
transformation. 
We also showed that the Lagrangean (\ref{eq3-1}) is invariant under the two valued
modular transformation. 
Lagrangean (\ref{eq3-1}) gives the masses of charged leptons and their super partners 
using electron mass as input. 
We use see-saw Lagrangean (\ref{eq3-2}) to obtain all the neutrino mass as input.
Similar prediction is done for charge + 2/3 quark masses, top quark mass as input and
charge -1/3 quark mass case, bottom quark mass as input. 
We use electron mass as input in the case of charged leptons.
Neutrino and super neutrino masses are determined: We use the heaviest
neutrino mass as the input. 
We use top quark mass and bottom quark mass in the quark mass
calculation as their inputs.

At the end, we emphasize that we can prove the supersymmetry using the
existing machines like b-factory or CERN LHC.

\appendix
\section{Angular distributions of Bhabha, $\mu \bar{\mu}$ pair and $\tilde{e}^+ \tilde{e}^-$ processes}
It is possible to measure the superelectron mass by comparing 
the angular distribution of the following processes.
\\
Bhabha scattering
\begin{equation}
\frac{d\sigma}{d\Omega}=\frac{\alpha^2}{4s}\Bigl(\frac{3 + \cos^2 \theta}{1 - \cos \theta}\Bigr)^{2}
\end{equation}
$\mu$-on pair production
\\
$e^+e^- \rightarrow \mu^{+} \mu^{-}$
\begin{equation}
\frac{d\sigma}{d\Omega}=\frac{\alpha^2 \beta_{\mu}}{4s}\Bigl(1 + \cos^2\theta + (1 - \beta^2_{\mu}\sin^2\theta)\Bigr)
\end{equation}
$e^+e^- \rightarrow \tilde{e}^+ \tilde{e}^-$
\begin{equation}
\frac{d\sigma}{d\Omega}= \frac{\alpha^2}{8s}\beta^3_{\tilde{e}} \sin^2 \theta
\end{equation}

\begin{figure}[h]
\begin{center}
\includegraphics[width=16cm]{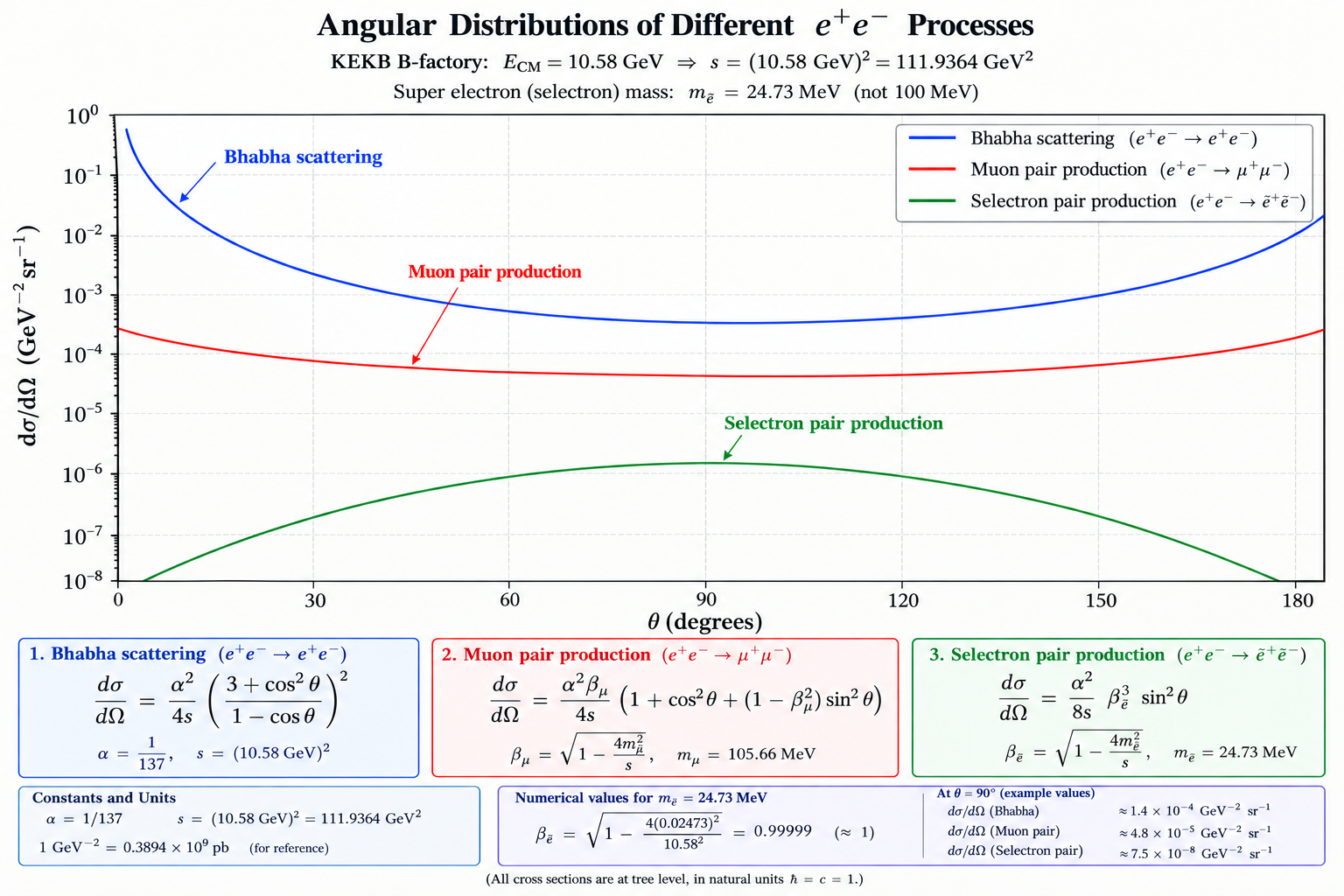}
\caption{}
\label{fig:1}
\end{center}
\end{figure}
%Please always give a title also for appendices.

\acknowledgments

We thank professor Toshiaki Kaneko and professor Fukuko Yuasa for their help in writing this paper.
We appreciate Director General professor Shoji Asai for pointing out the possible experiment to 
find the super electron using the b-factory beam dump.

%This is the most common positions for acknowledgments. A macro is
%available to maintain the same layout and spelling of the heading.

%\paragraph{Note added.} 

%This is also a good position for notes added
%after the paper has been written.

% The bibliography will probably be heavily edited during typesetting.
% We'll parse it and, using the arxiv number or the journal data, will
% query inspire, trying to verify the data (this will probalby spot
% eventual typos) and retrive the document DOI and eventual errata.
% We however suggest to always provide author, title and journal data:
% in short all the informations that clearly identify a document.


\begin{thebibliography}{99}

\bibitem{a}
S. Shiba Funai, H. Sugawara, \emph{Flavor Moonshine},
\emph{KEK-TH-2151};

S. Shiba Funai, H. Sugawara, \emph{Flavor Moonshine}, 
\emph{Prog. Theor. Exp. Phys.} (2020) 013 B03.

\bibitem{b}
H. Cohn and J. Deutsch, 
\emph{Application of Symbolic Manipulation to Hecke Transformations of
Modular Forms in Two Variables}, 
\emph{Mathematics of Computation} \textbf{48} No. 177 (1987) 139--146.

J.~Bagger and J.~Wess, \emph{Supersymmetry and Supergravity},
Princeton 1983.

\bibitem{c}
J. H. Conway and S. P. Norton, 
\emph{Monstrous Moonshine}, 
\emph{Bull. Lond. Math. Soc.} \textbf{11} (1979) 308--339,
DOI : 10.1112/blms/11.3.308.

%\bibitem{d}
%J. H. Conway, S. P. Norton, \emph{Monstrous Moonshine}, 
%\emph{Bulletin of the London Mathematical Society} 
% \textbf{11} (1979), No.3  308–339
%DOI : 10.1112/blms/11.3.308

\bibitem{e}
T. Eguchi, H. Ooguri and Y. Tachikawa, 
\emph{Notes on the K3 Surface and the Mathieu Group $M_{24}$}, 
\emph{Experimental mathematics} 
\textbf{20} (2011) 91--96. arXiv:1004.0956 [hep-th].

\bibitem{f}
H. Miyazawa and H. Sugawara, 
\emph{Symmetry SU (9)}, 
\emph{Prog. Theor. Phys.} \textbf{33} (1965) 771--772.

\bibitem{g}
H. Miyazawa and H. Sugawara, 
\emph{Intrisically broken ${\tilde{U}}$ (18) symmetry group}, 
\emph{Pro.Theo.Phys} \textbf{34} No. 2 (1965) 263--273.

\bibitem{h}
D.S. Akerib et al.,
\emph{Search for dark matter particle interactions in an extended nuclear recoil energy
window with the LUX-ZEPLIN (LZ) experiment}
\texttt{https://arXiv.org/pdf/2609.02823}

\bibitem{i}
N. Nagata and T. T. Yanagida
\emph{Asymmetric Inelastic Dark Matter and the LUX-ZEPLIN event}
\texttt{https://arXiv.org/pdf/2609.18564}

%\bibitem{b}
%Author, \emph{Title},
%arxiv:1234.5678.

%\bibitem{c}
%Author, \emph{Title},
%Publisher (year).

%\bibitem{a}
%Author, \emph{Title}, \emph{J. Abbrev.} \textbf{vol} (year) pg.
%
%\bibitem{b}
%Author, \emph{Title},
%arxiv:1234.5678.
%
%\bibitem{c}
%Author, \emph{Title},
%Publisher (year).


% Please avoid comments such as "For a review'', "For some examples",
% "and references therein" or move them in the text. In general,
% please leave only references in the bibliography and move all
% accessory text in footnotes.

% Also, please have only one work for each \bibitem.


\end{thebibliography}
\end{document}